\documentclass[11pt,nofonttune]{IEEEtran}
\usepackage{url}
\usepackage[utf8]{inputenc}
\usepackage{amsmath}
\usepackage{amssymb}
\usepackage{xspace}
\usepackage{epsfig}
\usepackage{balance}
\usepackage{listings}
\usepackage[dvipsnames]{xcolor}

\definecolor{codegray}{rgb}{0.25,0.25,0.25}
\definecolor{codepurple}{rgb}{0.58,0,0.82}
\lstdefinestyle{mystyle-yaml}{
  commentstyle=\color{gray},
  keywordstyle=\color{purple},
  numberstyle=\tiny\color{codegray},
  stringstyle=\color{codepurple},
  basicstyle=\color{Periwinkle}\ttfamily\scriptsize,
  rulecolor=\color{black},
  breakatwhitespace=true,         
  breaklines=true,                 
  captionpos=b,
  frame=tb,
  keepspaces=true,                 
  numbers=left,                    
  numbersep=5pt,                  
  showspaces=false,                
  showstringspaces=false,
  showtabs=false,                  
  tabsize=2,
  xleftmargin=10pt,
}

\lstdefinelanguage{yaml}{
  alsoletter={-},
  keywords={true,false,null,y,n,-},
  sensitive=false,
  comment=[l]{\#},
  morecomment=[s]{/*}{*/},
  moredelim=[l][\color{orange}]{\&},
  moredelim=[l][\color{magenta}]{*},
  moredelim=**[il][\color{purple}{:}\color{MidnightBlue}]{:},   
  morestring=[b]',
  morestring=[b]",
}

\usepackage[acronyms,nonumberlist,nopostdot,nomain,nogroupskip,acronymlists={hidden}]{glossaries}
\newglossary[algh]{hidden}{acrh}{acnh}{Hidden Acronyms}

\usepackage{booktabs}
\usepackage{tabularx}

\usepackage{tikz}
\usepackage{pgfplots}
\pgfplotsset{compat=newest}
\pgfplotsset{plot coordinates/math parser=false}
\newlength\fheight
\newlength\fwidth
\usetikzlibrary{plotmarks,patterns,decorations.pathreplacing,backgrounds,calc,arrows,arrows.meta,spy,matrix,scopes,pgfplots.colorbrewer}
\usepgfplotslibrary{patchplots,groupplots,colorbrewer}
\usepackage{tikzscale}
\usepackage[draft]{hyperref}

\newif\ifexttikz
\exttikzfalse

\ifexttikz
	\usetikzlibrary{external}
\fi

\usepackage{multirow}
\usepackage[font=footnotesize]{subcaption}
\usepackage[font=footnotesize]{caption}

\usepackage{mathtools}
\usepackage[numbers,sort&compress]{natbib}

\usepackage{soul}

\newacronym{3gpp}{3GPP}{3rd Generation Partnership Project}
\newacronym{4g}{4G}{4th generation}
\newacronym{5g}{5G}{5th generation}
\newacronym{6g}{6G}{6th generation}
\newacronym{5gc}{5GC}{5G Core}
\newacronym{adc}{ADC}{Analog to Digital Converter}
\newacronym{aerpaw}{AERPAW}{Aerial Experimentation and Research Platform for Advanced Wireless}
\newacronym{ai}{AI}{Artificial Intelligence}
\newacronym{aimd}{AIMD}{Additive Increase Multiplicative Decrease}
\newacronym{am}{AM}{Acknowledged Mode}
\newacronym{amc}{AMC}{Adaptive Modulation and Coding}
\newacronym{amf}{AMF}{Access and Mobility Management Function}
\newacronym{aops}{AOPS}{Adaptive Order Prediction Scheduling}
\newacronym{api}{API}{Application Programming Interface}
\newacronym{apn}{APN}{Access Point Name}
\newacronym{ap}{AP}{Application Protocol}
\newacronym{aqm}{AQM}{Active Queue Management}
\newacronym{ausf}{AUSF}{Authentication Server Function}
\newacronym{avc}{AVC}{Advanced Video Coding}
\newacronym{awgn}{AGWN}{Additive White Gaussian Noise}
\newacronym{balia}{BALIA}{Balanced Link Adaptation Algorithm}
\newacronym{bbu}{BBU}{Base Band Unit}
\newacronym{bdp}{BDP}{Bandwidth-Delay Product}
\newacronym{ber}{BER}{Bit Error Rate}
\newacronym{bf}{BF}{Beamforming}
\newacronym{bler}{BLER}{Block Error Rate}
\newacronym{brr}{BRR}{Bayesian Ridge Regressor}
\newacronym{bs}{BS}{Base Station}
\newacronym{bsr}{BSR}{Buffer Status Report}
\newacronym{bss}{BSS}{Business Support System}
\newacronym{ca}{CA}{Carrier Aggregation}
\newacronym{caas}{CaaS}{Connectivity-as-a-Service}
\newacronym{cb}{CB}{Code Block}
\newacronym{cc}{CC}{Congestion Control}
\newacronym{ccid}{CCID}{Congestion Control ID}
\newacronym{cco}{CC}{Carrier Component}
\newacronym{cd}{CD}{Continuous Delivery}
\newacronym{cdd}{CDD}{Cyclic Delay Diversity}
\newacronym{cdf}{CDF}{Cumulative Distribution Function}
\newacronym{cdn}{CDN}{Content Distribution Network}
\newacronym{ci}{CI}{Continuous Integration}
\newacronym{cli}{CLI}{Command-line Interface}
\newacronym{cn}{CN}{Core Network}
\newacronym{codel}{CoDel}{Controlled Delay Management}
\newacronym{comac}{COMAC}{Converged Multi-Access and Core}
\newacronym{cord}{CORD}{Central Office Re-architected as a Datacenter}
\newacronym{cornet}{CORNET}{COgnitive Radio NETwork}
\newacronym{cosmos}{COSMOS}{Cloud Enhanced Open Software Defined Mobile Wireless Testbed for City-Scale Deployment}
\newacronym{cots}{COTS}{Commercial Off-the-Shelf}
\newacronym{cp}{CP}{Control Plane}
\newacronym{cyp}{CP}{Cyclic Prefix}
\newacronym{up}{UP}{User Plane}
\newacronym{cpu}{CPU}{Central Processing Unit}
\newacronym{cqi}{CQI}{Channel Quality Information}
\newacronym{cr}{CR}{Cognitive Radio}
\newacronym{cran}{CRAN}{Cloud \gls{ran}}
\newacronym{crs}{CRS}{Cell Reference Signal}
\newacronym{csi}{CSI}{Channel State Information}
\newacronym{csirs}{CSI-RS}{Channel State Information - Reference Signal}
\newacronym{ct}{CT}{Continuous Testing}
\newacronym{cu}{CU}{Central Unit}
\newacronym{d2tcp}{D$^2$TCP}{Deadline-aware Data center TCP}
\newacronym{d3}{D$^3$}{Deadline-Driven Delivery}
\newacronym{dac}{DAC}{Digital to Analog Converter}
\newacronym{dag}{DAG}{Directed Acyclic Graph}
\newacronym{das}{DAS}{Distributed Antenna System}
\newacronym{dash}{DASH}{Dynamic Adaptive Streaming over HTTP}
\newacronym{dc}{DC}{Dual Connectivity}
\newacronym{dccp}{DCCP}{Datagram Congestion Control Protocol}
\newacronym{dce}{DCE}{Direct Code Execution}
\newacronym{dci}{DCI}{Downlink Control Information}
\newacronym{dctcp}{DCTCP}{Data Center TCP}
\newacronym{dl}{DL}{Downlink}
\newacronym{dmr}{DMR}{Deadline Miss Ratio}
\newacronym{dmrs}{DMRS}{DeModulation Reference Signal}
\newacronym{drlcc}{DRL-CC}{Deep Reinforcement Learning Congestion Control}
\newacronym{drs}{DRS}{Discovery Reference Signal}
\newacronym{du}{DU}{Distributed Unit}
\newacronym{e2e}{E2E}{end-to-end}
\newacronym{earfcn}{EARFCN}{E-UTRA Absolute Radio Frequency Channel Number}
\newacronym{ecaas}{ECaaS}{Edge-Cloud-as-a-Service}
\newacronym{ecn}{ECN}{Explicit Congestion Notification}
\newacronym{edf}{EDF}{Earliest Deadline First}
\newacronym{embb}{eMBB}{Enhanced Mobile Broadband}
\newacronym{empower}{EMPOWER}{EMpowering transatlantic PlatfOrms for advanced WirEless Research}
\newacronym{enb}{eNB}{evolved Node Base}
\newacronym{endc}{EN-DC}{E-UTRAN-\gls{nr} \gls{dc}}
\newacronym{epc}{EPC}{Evolved Packet Core}
\newacronym{eps}{EPS}{Evolved Packet System}
\newacronym{es}{ES}{Edge Server}
\newacronym{etsi}{ETSI}{European Telecommunications Standards Institute}
\newacronym[firstplural=Estimated Times of Arrival (ETAs)]{eta}{ETA}{Estimated Time of Arrival}
\newacronym{eutran}{E-UTRAN}{Evolved Universal Terrestrial Access Network}
\newacronym{faas}{FaaS}{Function-as-a-Service}
\newacronym{fapi}{FAPI}{Functional Application Platform Interface}
\newacronym{fdd}{FDD}{Frequency Division Duplexing}
\newacronym{fdm}{FDM}{Frequency Division Multiplexing}
\newacronym{fdma}{FDMA}{Frequency Division Multiple Access}
\newacronym{fed4fire}{FED4FIRE+}{Federation 4 Future Internet Research and Experimentation Plus}
\newacronym{fir}{FIR}{Finite Impulse Response}
\newacronym{fit}{FIT}{Future \acrlong{iot}}
\newacronym{fpga}{FPGA}{Field Programmable Gate Array}
\newacronym{fr2}{FR2}{Frequency Range 2}
\newacronym{fs}{FS}{Fast Switching}
\newacronym{fscc}{FSCC}{Flow Sharing Congestion Control}
\newacronym{ftp}{FTP}{File Transfer Protocol}
\newacronym{fw}{FW}{Flow Window}
\newacronym{ge}{GE}{Gaussian Elimination}
\newacronym{gnb}{gNB}{Next Generation Node Base}
\newacronym{gop}{GOP}{Group of Pictures}
\newacronym{gpr}{GPR}{Gaussian Process Regressor}
\newacronym{gpu}{GPU}{Graphics Processing Unit}
\newacronym{gtp}{GTP}{GPRS Tunneling Protocol}
\newacronym{gtpc}{GTP-C}{GPRS Tunnelling Protocol Control Plane}
\newacronym{gtpu}{GTP-U}{GPRS Tunnelling Protocol User Plane}
\newacronym{gtpv2c}{GTPv2-C}{\gls{gtp} v2 - Control}
\newacronym{gw}{GW}{Gateway}
\newacronym{harq}{HARQ}{Hybrid Automatic Repeat reQuest}
\newacronym{hetnet}{HetNet}{Heterogeneous Network}
\newacronym{hh}{HH}{Hard Handover}
\newacronym{hol}{HOL}{Head-of-Line}
\newacronym{hqf}{HQF}{Highest-quality-first}
\newacronym{hss}{HSS}{Home Subscription Server}
\newacronym{http}{HTTP}{HyperText Transfer Protocol}
\newacronym{ia}{IA}{Initial Access}
\newacronym{iab}{IAB}{Integrated Access and Backhaul}
\newacronym{ic}{IC}{Incident Command}
\newacronym{ietf}{IETF}{Internet Engineering Task Force}
\newacronym{imsi}{IMSI}{International Mobile Subscriber Identity}
\newacronym{imt}{IMT}{International Mobile Telecommunication}
\newacronym{iot}{IoT}{Internet of Things}
\newacronym{ip}{IP}{Internet Protocol}
\newacronym{itu}{ITU}{International Telecommunication Union}
\newacronym{k8s}{k8s}{Kubernetes}
\newacronym{kpi}{KPI}{Key Performance Indicator}
\newacronym{kpm}{KPM}{Key Performance Measurement}
\newacronym{kvm}{KVM}{Kernel-based Virtual Machine}
\newacronym{los}{LOS}{Line-of-Sight}
\newacronym{lsm}{LSM}{Link-to-System Mapping}
\newacronym{lstm}{LSTM}{Long Short Term Memory}
\newacronym{lte}{LTE}{Long Term Evolution}
\newacronym{lxc}{LXC}{Linux Container}
\newacronym{m2m}{M2M}{Machine to Machine}
\newacronym{mac}{MAC}{Medium Access Control}
\newacronym{manet}{MANET}{Mobile Ad Hoc Network}
\newacronym{mano}{MANO}{Management and Orchestration}
\newacronym{mc}{MC}{Multi-Connectivity}
\newacronym{mcc}{MCC}{Mobile Cloud Computing}
\newacronym{mchem}{MCHEM}{Massive Channel Emulator}
\newacronym{mcs}{MCS}{Modulation and Coding Scheme}
\newacronym{mec}{MEC}{Multi-access Edge Computing}
\newacronym{mec2}{MEC}{Mobile Edge Cloud}
\newacronym{mfc}{MFC}{Mobile Fog Computing}
\newacronym{mgen}{MGEN}{Multi-Generator}
\newacronym{mi}{MI}{Mutual Information}
\newacronym{mib}{MIB}{Master Information Block}
\newacronym{miesm}{MIESM}{Mutual Information Based Effective SINR}
\newacronym{mimo}{MIMO}{Multiple Input, Multiple Output}
\newacronym{ml}{ML}{Machine Learning}
\newacronym{mlr}{MLR}{Maximum-local-rate}
\newacronym[plural=\gls{mme}s,firstplural=Mobility Management Entities (MMEs)]{mme}{MME}{Mobility Management Entity}
\newacronym{mmtc}{mMTC}{Massive Machine-Type Communications}
\newacronym{mmwave}{mmWave}{millimeter wave}
\newacronym{mno}{MNO}{Mobile Network Operator}
\newacronym{mpdccp}{MP-DCCP}{Multipath Datagram Congestion Control Protocol}
\newacronym{mptcp}{MPTCP}{Multipath TCP}
\newacronym{mr}{MR}{Maximum Rate}
\newacronym{mrdc}{MR-DC}{Multi \gls{rat} \gls{dc}}
\newacronym{mse}{MSE}{Mean Square Error}
\newacronym{mss}{MSS}{Maximum Segment Size}
\newacronym{mt}{MT}{Mobile Termination}
\newacronym{mtd}{MTD}{Machine-Type Device}
\newacronym{mtu}{MTU}{Maximum Transmission Unit}
\newacronym{mumimo}{MU-MIMO}{Multi-user \gls{mimo}}
\newacronym{mvno}{MVNO}{Mobile Virtual Network Operator}
\newacronym{nalu}{NALU}{Network Abstraction Layer Unit}
\newacronym{nas}{NAS}{Network Attached Storage}
\newacronym{nat}{NAT}{Network Address Translation}
\newacronym{nbiot}{NB-IoT}{Narrow Band IoT}
\newacronym{nfs}{NFS}{Network File System}
\newacronym{nfv}{NFV}{Network Function Virtualization}
\newacronym{nfvi}{NFVI}{Network Function Virtualization Infrastructure}
\newacronym{ni}{NI}{Network Interfaces}
\newacronym{nic}{NIC}{Network Interface Card}
\newacronym{nlos}{NLOS}{Non-Line-of-Sight}
\newacronym{now}{NOW}{Non Overlapping Window}
\newacronym{nsm}{NSM}{Network Service Mesh}
\newacronym[type=hidden]{nr}{NR}{New Radio}
\newacronym{nrf}{NRF}{Network Repository Function}
\newacronym{nsa}{NSA}{Non Stand Alone}
\newacronym{nse}{NSE}{Network Slicing Engine}
\newacronym{nssf}{NSSF}{Network Slice Selection Function}
\newacronym{o2i}{O2I}{Outdoor to Indoor}
\newacronym{oai}{OAI}{OpenAirInterface}
\newacronym{oaicn}{OAI-CN}{\gls{oai} \acrlong{cn}}
\newacronym{oairan}{OAI-RAN}{\acrlong{oai} \acrlong{ran}}
\newacronym{oam}{OAM}{Operations, Administration and Maintenance}
\newacronym{oci}{OCI}{Open Container Initiative}
\newacronym{ofdm}{OFDM}{Orthogonal Frequency Division Multiplexing}
\newacronym{olia}{OLIA}{Opportunistic Linked Increase Algorithm}
\newacronym{omec}{OMEC}{Open Mobile Evolved Core}
\newacronym{onap}{ONAP}{Open Network Automation Platform}
\newacronym{onf}{ONF}{Open Networking Foundation}
\newacronym{onos}{ONOS}{Open Networking Operating System}
\newacronym{oom}{OOM}{\gls{onap} Operations Manager}
\newacronym{opnfv}{OPNFV}{Open Platform for \gls{nfv}}
\newacronym[type=hidden]{oran}{O-RAN}{Open \gls{ran}}
\newacronym{orbit}{ORBIT}{Open-Access Research Testbed for Next-Generation Wireless Networks}
\newacronym{os}{OS}{Operating System}
\newacronym{oss}{OSS}{Operations Support System}
\newacronym{pa}{PA}{Position-aware}
\newacronym{pase}{PASE}{Prioritization, Arbitration, and Self-adjusting Endpoints}
\newacronym{pawr}{PAWR}{Platforms for Advanced Wireless Research}
\newacronym{pbch}{PBCH}{Physical Broadcast Channel}
\newacronym{pcef}{PCEF}{Policy and Charging Enforcement Function}
\newacronym{pcfich}{PCFICH}{Physical Control Format Indicator Channel}
\newacronym{pcrf}{PCRF}{Policy and Charging Rules Function}
\newacronym{pdcch}{PDCCH}{Physical Downlink Control Channel}
\newacronym{pdcp}{PDCP}{Packet Data Convergence Protocol}
\newacronym{pdsch}{PDSCH}{Physical Downlink Shared Channel}
\newacronym{pdu}{PDU}{Packet Data Unit}
\newacronym{pf}{PF}{Proportional Fair}
\newacronym{pgw}{PGW}{Packet Gateway}
\newacronym{phich}{PHICH}{Physical Hybrid ARQ Indicator Channel}
\newacronym{phy}{PHY}{Physical}
\newacronym{pmch}{PMCH}{Physical Multicast Channel}
\newacronym{pmi}{PMI}{Precoding Matrix Indicators}
\newacronym{powder}{POWDER}{Platform for Open Wireless Data-driven Experimental Research}
\newacronym{ppo}{PPO}{Proximal Policy Optimization}
\newacronym{ppp}{PPP}{Poisson Point Process}
\newacronym{prach}{PRACH}{Physical Random Access Channel}
\newacronym{prb}{PRB}{Physical Resource Block}
\newacronym{psnr}{PSNR}{Peak Signal to Noise Ratio}
\newacronym{pss}{PSS}{Primary Synchronization Signal}
\newacronym{pucch}{PUCCH}{Physical Uplink Control Channel}
\newacronym{pusch}{PUSCH}{Physical Uplink Shared Channel}
\newacronym{qam}{QAM}{Quadrature Amplitude Modulation}
\newacronym{qci}{QCI}{\gls{qos} Class Identifier}
\newacronym{qoe}{QoE}{Quality of Experience}
\newacronym{qos}{QoS}{Quality of Service}
\newacronym{quic}{QUIC}{Quick UDP Internet Connections}
\newacronym{rach}{RACH}{Random Access Channel}
\newacronym{ran}{RAN}{Radio Access Network}
\newacronym[firstplural=Radio Access Technologies (RATs)]{rat}{RAT}{Radio Access Technology}
\newacronym{rbg}{RBG}{Resource Block Group}
\newacronym{rcn}{RCN}{Research Coordination Network}
\newacronym{rc}{RC}{RAN Control}
\newacronym{rec}{REC}{Radio Edge Cloud}
\newacronym{red}{RED}{Random Early Detection}
\newacronym{renew}{RENEW}{Reconfigurable Eco-system for Next-generation End-to-end Wireless}
\newacronym{rf}{RF}{Radio Frequency}
\newacronym{rfc}{RFC}{Request for Comments}
\newacronym{rfr}{RFR}{Random Forest Regressor}
\newacronym{ric}{RIC}{\gls{ran} Intelligent Controller}
\newacronym{rlc}{RLC}{Radio Link Control}
\newacronym{rlf}{RLF}{Radio Link Failure}
\newacronym{rlnc}{RLNC}{Random Linear Network Coding}
\newacronym{rmr}{RMR}{RIC Message Router}
\newacronym{rmse}{RMSE}{Root Mean Squared Error}
\newacronym{rnis}{RNIS}{Radio Network Information Service}
\newacronym{rr}{RR}{Round Robin}
\newacronym{rrc}{RRC}{Radio Resource Control}
\newacronym{rrm}{RRM}{Radio Resource Management}
\newacronym{rru}{RRU}{Remote Radio Unit}
\newacronym{rs}{RS}{Remote Server}
\newacronym{rsrp}{RSRP}{Reference Signal Received Power}
\newacronym{rsrq}{RSRQ}{Reference Signal Received Quality}
\newacronym{rss}{RSS}{Received Signal Strength}
\newacronym{rssi}{RSSI}{Received Signal Strength Indicator}
\newacronym{rtt}{RTT}{Round Trip Time}
\newacronym{ru}{RU}{Radio Unit}
\newacronym{rw}{RW}{Receive Window}
\newacronym{rx}{RX}{Receiver}
\newacronym{s1ap}{S1AP}{S1 Application Protocol}
\newacronym{sa}{SA}{standalone}
\newacronym{sack}{SACK}{Selective Acknowledgment}
\newacronym{sap}{SAP}{Service Access Point}
\newacronym{sc2}{SC2}{Spectrum Collaboration Challenge}
\newacronym{scef}{SCEF}{Service Capability Exposure Function}
\newacronym{sch}{SCH}{Secondary Cell Handover}
\newacronym{scoot}{SCOOT}{Split Cycle Offset Optimization Technique}
\newacronym{sctp}{SCTP}{Stream Control Transmission Protocol}
\newacronym{sdap}{SDAP}{Service Data Adaptation Protocol}
\newacronym{sdk}{SDK}{Software Development Kit}
\newacronym{sdm}{SDM}{Space Division Multiplexing}
\newacronym{sdma}{SDMA}{Spatial Division Multiple Access}
\newacronym{sdn}{SDN}{Software-defined Networking}
\newacronym{sdr}{SDR}{Software-defined Radio}
\newacronym{seba}{SEBA}{SDN-Enabled Broadband Access}
\newacronym{sgsn}{SGSN}{Serving GPRS Support Node}
\newacronym{sgw}{SGW}{Service Gateway}
\newacronym{si}{SI}{Study Item}
\newacronym{sib}{SIB}{Secondary Information Block}
\newacronym{sinr}{SINR}{Signal to Interference plus Noise Ratio}
\newacronym{sip}{SIP}{Session Initiation Protocol}
\newacronym{siso}{SISO}{Single Input, Single Output}
\newacronym{sla}{SLA}{Service Level Agreement}
\newacronym{sm}{SM}{Service Model}
\newacronym{smf}{SMF}{Session Management Function}
\newacronym{smo}{SMO}{Service Management and Orchestration}
\newacronym{sms}{SMS}{Short Message Service}
\newacronym{smsgmsc}{SMS-GMSC}{\gls{sms}-Gateway}
\newacronym{snr}{SNR}{Signal-to-Noise-Ratio}
\newacronym{son}{SON}{Self-Organizing Network}
\newacronym{sptcp}{SPTCP}{Single Path TCP}
\newacronym{srb}{SRB}{Service Radio Bearer}
\newacronym{sriov}{SR-IOV}{Single Root Input/Output Virtualization}
\newacronym{srn}{SRN}{Standard Radio Node}
\newacronym{srs}{SRS}{Sounding Reference Signal}
\newacronym{ss}{SS}{Synchronization Signal}
\newacronym{sss}{SSS}{Secondary Synchronization Signal}
\newacronym{st}{ST}{Spanning Tree}
\newacronym{svc}{SVC}{Scalable Video Coding}
\newacronym{tb}{TB}{Transport Block}
\newacronym{tcp}{TCP}{Transmission Control Protocol}
\newacronym{tdd}{TDD}{Time Division Duplexing}
\newacronym{tdm}{TDM}{Time Division Multiplexing}
\newacronym{tdma}{TDMA}{Time Division Multiple Access}
\newacronym{tfl}{TfL}{Transport for London}
\newacronym{tfrc}{TFRC}{TCP-Friendly Rate Control}
\newacronym{tft}{TFT}{Traffic Flow Template}
\newacronym{tgen}{TGEN}{Traffic Generator}
\newacronym{tip}{TIP}{Telecom Infra Project}
\newacronym{tm}{TM}{Transparent Mode}
\newacronym{to}{TO}{Telco Operator}
\newacronym{tr}{TR}{Technical Report}
\newacronym{trp}{TRP}{Transmitter Receiver Pair}
\newacronym{ts}{TS}{Technical Specification}
\newacronym{tti}{TTI}{Transmission Time Interval}
\newacronym{ttt}{TTT}{Time-to-Trigger}
\newacronym{tx}{TX}{Transmitter}
\newacronym{uas}{UAS}{Unmanned Aerial System}
\newacronym{uav}{UAV}{Unmanned Aerial Vehicle}
\newacronym{udm}{UDM}{Unified Data Management}
\newacronym{udp}{UDP}{User Datagram Protocol}
\newacronym{udr}{UDR}{Unified Data Repository}
\newacronym{ue}{UE}{User Equipment}
\newacronym{uhd}{UHD}{\gls{usrp} Hardware Driver}
\newacronym{ul}{UL}{Uplink}
\newacronym{um}{UM}{Unacknowledged Mode}
\newacronym{uml}{UML}{Unified Modeling Language}
\newacronym{upa}{UPA}{Uniform Planar Array}
\newacronym{upf}{UPF}{User Plane Function}
\newacronym{urllc}{URLLC}{Ultra Reliable and Low Latency Communications}
\newacronym{usa}{U.S.}{United States}
\newacronym{usim}{USIM}{Universal Subscriber Identity Module}
\newacronym{usrp}{USRP}{Universal Software Radio Peripheral}
\newacronym{utc}{UTC}{Urban Traffic Control}
\newacronym{vim}{VIM}{Virtualization Infrastructure Manager}
\newacronym{vm}{VM}{Virtual Machine}
\newacronym{vnf}{VNF}{Virtual Network Function}
\newacronym{volte}{VoLTE}{Voice over \gls{lte}}
\newacronym{voltha}{VOLTHA}{Virtual OLT HArdware Abstraction}
\newacronym{vr}{VR}{Virtual Reality}
\newacronym{vran}{vRAN}{Virtualized \gls{ran}}
\newacronym{vss}{VSS}{Video Streaming Server}
\newacronym{wbf}{WBF}{Wired Bias Function}
\newacronym{wf}{WF}{Waterfilling}
\newacronym{wg}{WG}{Working Group}
\newacronym{wlan}{WLAN}{Wireless Local Area Network}
\newacronym{osm}{OSM}{Open Source \gls{nfv} Management and Orchestration}
\newacronym{pnf}{PNF}{Physical Network Function}
\newacronym{drl}{DRL}{Deep Reinforcement Learning}
\newacronym{mtc}{MTC}{Machine-type Communications}
\newacronym{osc}{OSC}{O-RAN Software Community}
\newacronym{mns}{MnS}{Management Services}
\newacronym{ves}{VES}{\gls{vnf} Event Stream}
\newacronym{ei}{EI}{Enrichment Information}
\newacronym{fh}{FH}{Fronthaul}
\newacronym{fft}{FFT}{Fast Fourier Transform}
\newacronym{laa}{LAA}{Licensed-Assisted Access}
\newacronym{plfs}{PLFS}{Physical Layer Frequency Signals}
\newacronym{ptp}{PTP}{Precision Time Protocol}
\newacronym{cbrs}{CBRS}{Citizen Broadband Radio Service}
\tikzstyle{startstop} = [rectangle, rounded corners, minimum width=2cm, minimum height=0.5cm,text centered, draw=black]
\tikzstyle{io} = [trapezium, trapezium left angle=70, trapezium right angle=110, minimum width=3cm, minimum height=1cm, text centered, draw=black]
\tikzstyle{process} = [rectangle, minimum width=2cm, minimum height=0.5cm, text centered, draw=black, alignb=center]
\tikzstyle{decision} = [ellipse, minimum width=2cm, minimum height=1cm, text centered, draw=black]
\tikzstyle{arrow} = [thick,<->,>=stealth]
\tikzstyle{line} = [thick,>=stealth]
\tikzstyle{darrow} = [thick,<->,>=stealth,dashed]
\tikzstyle{sarrow} = [thick,->,>=stealth]
\tikzstyle{larrow} = [line width=0.1mm,dashdotted,->,>=stealth]
\tikzstyle{llarrow} = [line width=0.1mm,->,>=stealth]
\tikzstyle{doublearrow} = [line width=0.1mm,<->,>=stealth]

\makeatletter
\def\grd@save@target#1{%
  \def\grd@target{#1}}
\def\grd@save@start#1{%
  \def\grd@start{#1}}
\tikzset{
  grid with coordinates/.style={
    to path={%
      \pgfextra{%
        \edef\grd@@target{(\tikztotarget)}%
        \tikz@scan@one@point\grd@save@target\grd@@target\relax
        \edef\grd@@start{(\tikztostart)}%
        \tikz@scan@one@point\grd@save@start\grd@@start\relax
        \draw[minor help lines] (\tikztostart) grid (\tikztotarget);
        \draw[major help lines] (\tikztostart) grid (\tikztotarget);
        \grd@start
        \pgfmathsetmacro{\grd@xa}{\the\pgf@x/1cm}
        \pgfmathsetmacro{\grd@ya}{\the\pgf@y/1cm}
        \grd@target
        \pgfmathsetmacro{\grd@xb}{\the\pgf@x/1cm}
        \pgfmathsetmacro{\grd@yb}{\the\pgf@y/1cm}
        \pgfmathsetmacro{\grd@xc}{\grd@xa + \pgfkeysvalueof{/tikz/grid with coordinates/major step x}}
        \pgfmathsetmacro{\grd@yc}{\grd@ya + \pgfkeysvalueof{/tikz/grid with coordinates/major step y}}
        \foreach \x in {\grd@xa,\grd@xc,...,\grd@xb}
        \node[anchor=north] at (\x,\grd@ya) {\pgfmathprintnumber{\x}};
        \foreach \y in {\grd@ya,\grd@yc,...,\grd@yb}
        \node[anchor=east] at (\grd@xa,\y) {\pgfmathprintnumber{\y}};
      }
    }
  },
  minor help lines/.style={
    help lines,
    gray,
    line cap =round,
    xstep=\pgfkeysvalueof{/tikz/grid with coordinates/minor step x},
    ystep=\pgfkeysvalueof{/tikz/grid with coordinates/minor step y}
  },
  major help lines/.style={
    help lines,
    line cap =round,
    line width=\pgfkeysvalueof{/tikz/grid with coordinates/major line width},
    xstep=\pgfkeysvalueof{/tikz/grid with coordinates/major step x},
    ystep=\pgfkeysvalueof{/tikz/grid with coordinates/major step y}
  },
  grid with coordinates/.cd,
  minor step x/.initial=.5,
  minor step y/.initial=.2,
  major step x/.initial=1,
  major step y/.initial=1,
  major line width/.initial=1pt,
}
\makeatother

\definecolor{desireRed}{RGB}{230,57,60}%
\definecolor{darkPurple}{RGB}{59,31,43}%
\definecolor{springGreen}{RGB}{37,223,145}%
\definecolor{queenBlue}{RGB}{69,123,157}%
\definecolor{spaceCadet}{RGB}{29,53,87}%

\usepackage{dblfloatfix}

\newcommand{\openran}{Open \gls{ran}\xspace}
\newcommand{\cicd}{\gls{ci}/\gls{cd}\xspace}
\newcommand{\cicdct}{\gls{ci}/\gls{cd}/\gls{ct}\xspace}

\newcommand{\ran}{\gls{ran}\xspace}
\newcommand{\ric}{\gls{ric}\xspace}

\newcommand{\name}{5G-CT\xspace}

\usepackage[var0]{inconsolata}

\ifexttikz
\else
\usepackage{tikzpagenodes,etoolbox}
\usetikzlibrary{calc}
\usepackage[contents={}]{background}
\AddEverypageHook{%
\ifnumequal{\thepage}{1}{%
    \tikz[remember picture,overlay]{%
        \node[draw,
        minimum width=1.03\textwidth,
        text width=1.02\textwidth,
        font=\scriptsize
        ]
        at ($(current page header area) - (0,5pt)$)
        {%
        This paper has been accepted for publication on IEEE Communications Magazine. This is the author's accepted version of the article. The final version published by IEEE is 
        L. Bonati, M. Polese, S. D’Oro, P. Brach del Prever, and T. Melodia, ``\name: Automated Deployment and Over-the-Air Testing of End-to-End Open Radio Access Networks,'' \textit{IEEE Communications Magazine,} 2024.
        };
        \node[draw,
        minimum width=1.03\textwidth,
        text width=1.02\textwidth,
        font=\scriptsize
        ]
        at ($(current page footer area) - (0,4pt)$)
        {%
        ©2024 IEEE. Personal use of this material is permitted. Permission from IEEE must be obtained for all other uses, in any current or future media, including reprinting/republishing this material for advertising or promotional purposes, creating new collective works, for resale or redistribution to servers or lists, or reuse of any copyrighted component of this work in other works.
        };
    }%
}{}
}
\fi

\begin{document}

\title{\name: Automated Deployment and Over-the-Air Testing of End-to-End Open Radio Access Networks}

\author{\IEEEauthorblockN{Leonardo Bonati, Michele Polese, Salvatore D'Oro, Pietro Brach del Prever, Tommaso Melodia}
\thanks{The authors are with the Institute for the Wireless Internet of Things, Northeastern University, Boston, MA, USA. E-mail: \{l.bonati, m.polese, s.doro, brachdelprever.p, melodia\}@northeastern.edu.}
\thanks{This article is based upon work partially supported by the U.S.\ National Science Foundation under grants CNS-1925601, CNS-2112471, and CNS-1923789, by the U.S.\ Office of Naval Research under grant N00014-20-1-2132, and by OUSD(R\&E) through Army Research Laboratory Cooperative Agreement Number W911NF-19-2-0221. The views and conclusions contained in this document are those of the authors and should not be interpreted as representing the official policies, either expressed or implied, of the Army Research Laboratory or the U.S. Government. The U.S. Government is authorized to reproduce and distribute reprints for Government purposes notwithstanding any copyright notation herein.}
}

\flushbottom
\setlength{\parskip}{0ex plus0.1ex}

\maketitle
\glsunset{nr}
\glsunset{lte}
\glsunset{3gpp}
\glsunset{cbrs}
\glsunset{udp}

\begin{abstract}

Deploying and testing cellular networks is a complex task due to the multitude of components involved---from the core to the \gls{ran} and \gls{ue}---all of which requires integration and constant monitoring.
Additional challenges are posed by the nature of the wireless channel, whose inherent randomness hinders the repeatability and consistency of the testing process.
Consequently, existing solutions for both private and public cellular systems still rely heavily on human intervention for operations such as network reconfiguration, performance monitoring, and end-to-end testing. This reliance significantly slows the pace of innovation in cellular systems.
To address these challenges, we introduce \name, an automation framework based on OpenShift and the GitOps workflow, capable of deploying a softwarized end-to-end 5G and O-RAN-compliant system in a matter of seconds without the need for any human intervention. We have deployed \name to test the integration and performance of open-source cellular stacks, including \acrlong{oai}, and have collected months of automated over-the-air testing results involving software-defined radios.
\name brings cloud-native continuous integration and delivery to the \gls{ran}, effectively addressing the complexities associated with managing spectrum, radios, heterogeneous devices, and distributed components.%
Moreover, it endows cellular networks with much needed automation and continuous testing capabilities, providing a platform to evaluate the robustness and resiliency of Open \gls{ran} software.

\end{abstract}

\begin{IEEEkeywords}
Open RAN, Automation, GitOps, 5G, 6G.\vspace{-.5cm}
\end{IEEEkeywords}

\glsresetall
\glsunset{nr}
\glsunset{lte}
\glsunset{3gpp}
\glsunset{cbrs}
\glsunset{udp}

\section{Introduction}

The Open \gls{ran} architecture developed by the O-RAN Alliance, as well as the evolution in \gls{3gpp} LTE and NR designs, are moving cellular networks toward disaggregated, softwarized, programmable intelligent systems~\cite{polese2023understanding}. \gls{ran} disaggregation and softwarization can break the current vendor lock-in and open the cellular ecosystem to a larger number of players, facilitating innovation in the cellular market.

\begin{figure*}[t]
    \centering
    \includegraphics[width=0.9\textwidth]{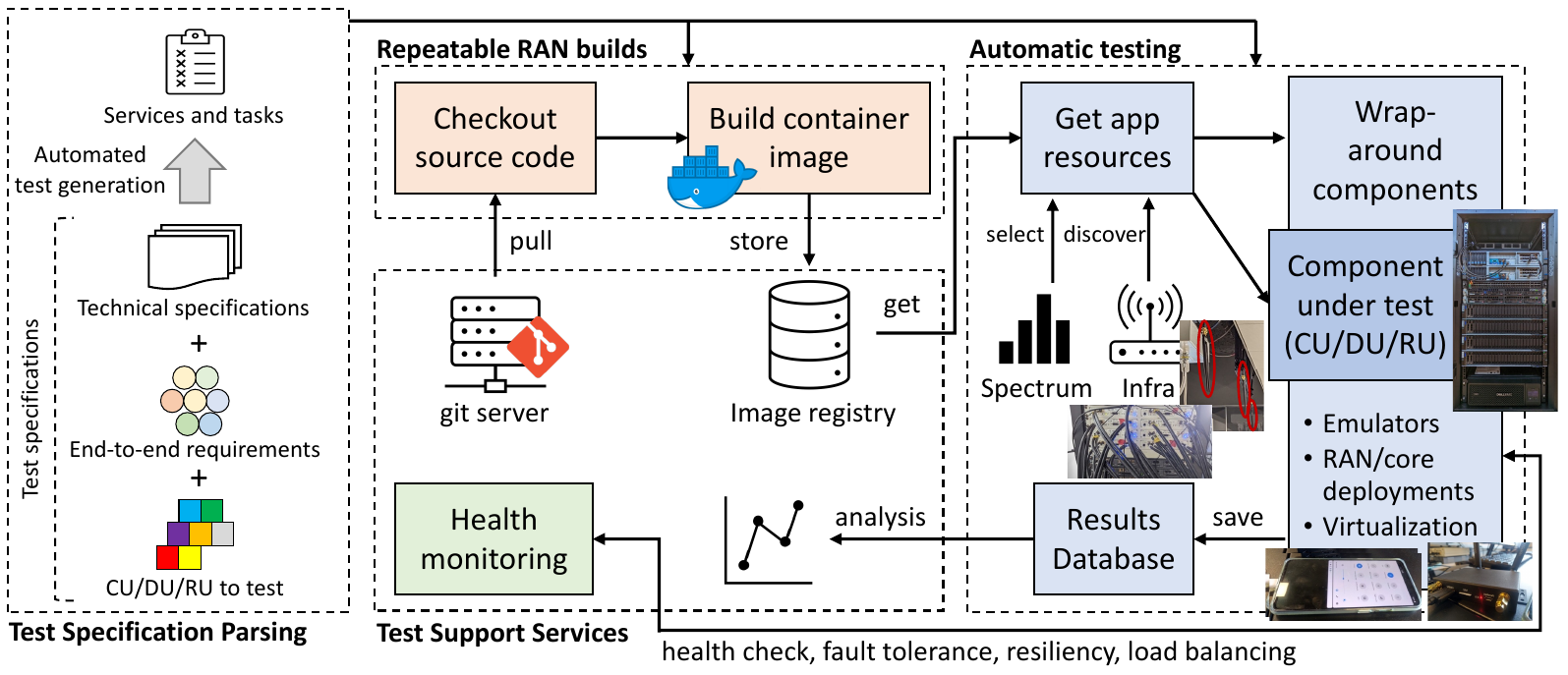}
    \caption{Intent-based CI/CD/CT pipeline for Open RAN.}
    \label{fig:pipeline-diagram}
\end{figure*}

However, network disaggregation and multi-vendor deployments are a double-edged sword: they provide much needed pathways to market diversity and supply chain robustness, but they also introduce \textit{interoperability, performance, and security challenges.} Interoperability testing can decrease the risks associated with Open \gls{ran} deployments and help to achieve feature and performance parity with traditional, inflexible systems.
Cellular deployment and testing are challenging as they are mostly manual and require significant human intervention to install, configure, update, monitor, and evaluate network components.

This issue affects both public and private networks. Public networks become inflexible, costly, and difficult to update or upgrade. Consequently, they often lag behind the rapid pace of innovation in the wireless domain and may harbor unpatched vulnerabilities. In the case of private networks, the complexity of end-to-end cellular systems necessitates specialized skills that are not typically found within the enterprise workforce. This makes the deployment of high-performance, robust cellular systems more costly compared to unlicensed technologies~\cite{chenumolu2023open}.
Therefore, it is clear that integrating automation in wireless networks has to go beyond software and needs to also span into radio components, distributed and heterogeneous devices, and spectrum bands.

DevOps, a portmanteau of the words software \emph{de\-v}el\-op\-ment and infrastructure \emph{op}\-er\-a\-tion\emph{s}, can be used to streamline the integration, deployment, and testing of code on the compute infrastructure~\cite{rodriguez2017continuous}.
This paradigm, also called \textit{infrastructure as code}, provides automation and tracking to ensure reliable and rapid delivery of new code and functionalities, while maintaining an authoritative source for the system and infrastructure configuration.

In this paper, we take a fundamental step toward cloud-native automation for the deployment and testing of open, programmable, multivendor end-to-end cellular networks. Specifically, we design, prototype, and evaluate \name, a set of automated pipelines, microservices, and infrastructure that can deploy a complete end-to-end 5G- and O-RAN-compliant cellular network in a few tens of seconds. \name leverages Red Hat OpenShift orchestration on a compute cluster which supports multiple core networks, edge, \gls{ran} (including the radio frontends), and \glspl{ric} deployed as microservices. Additional microservices support the automation and the management of radios and spectrum. GitOps~\cite{scotece20235g} pipelines---a specific class of DevOps based on \texttt{git} repositories---implement \gls{ci} and \gls{cd} to track authoritative sources for the codebase and the infrastructure configuration, and keep the deployment environment synchronized and updated.

\name is deployed and operated over-the-air in the Northeastern University FCC Innovation Zone. For months, it has been collecting test results (discussed later in this paper) to evaluate the performance of an end-to-end cellular network that includes the \gls{oai} 5G stack~\cite{kaltenberger2019openairinterface}, commercial 5G \glspl{ue}, and multiple core networks, i.e., the open-source Open5GS and a commercial core from A5G Networks.
By leveraging such automation, \name can effectively guarantee that heterogeneous, multivendor components can interoperate in a disaggregated and virtualized Open \gls{ran} system, and that code updates do not introduce regressions, while monitoring the system performance.

\section{Related Work}

DevOps techniques are widely used in the cloud computing domain. For instance, \cite{singh2019comparison} showcases a cloud-based \cicd workflow to build Docker images from code on remote repositories.
However, this work is neither concerned with the automated over-the-air testing of \gls{ran} components, nor with evaluating how updated builds affect \gls{ran} \glspl{kpi}.

Adopting DevOps techniques in cellular systems is challenging because of the heterogeneity of infrastructure, code, and functionalities in wireless networks. Moreover, DevOps for cellular needs to manage spectrum resources and guarantee predetermined \gls{qos} levels to the end users.
So far, the literature has mostly focused on the challenges to efficiently transition \gls{ran} workloads into microservices~\cite{schmidt2019flexvran,moorthy2022oswireless,foukas2017orion,garciaaviles2021nuberu}. Even though these solutions provide enhanced and automated network control, they either do not consider the challenges involving the automated instantiation of the \gls{ran}, or the automated testing of \gls{ran} code and functionalities over the air, as we do in this work.
Technologies to enable \cicd and automated instantiation of \gls{ran} components are discussed in~\cite{chenumolu2023open}, which also provides insights on how to fine-tune the compute machines where \gls{ran} functions are deployed. Differently from our paper, this work does not focus on the actual prototyping of the described \cicd and is not concerned with testing automation.

The \gls{oai} project has developed and maintains a \gls{ci} framework to run integration and testing for the \ran and core network components of the project, including 4G and 5G versions. While this toolchain contributes to the quality assurance process for \gls{oai}, it is not focused on deploying an end-to-end network on a production environment, and the radio testing is performed within a confined environment (i.e., a small Faraday cage)~\cite{defosseux2019oai}. The authors of~\cite{arouk2020kube} integrate \gls{oai} for a \gls{k8s}-based \gls{cd} framework, which however does not embed CT capabilities. Other works leverage DevOps for \ran slicing~\cite{li2021automated} or core network management~\cite{scotece20235g}, but do not consider the end-to-end \ran, core, and edge services (e.g., the \ric) deployment, as we do in this paper.

\section{Automated Open RAN Pipelines}
\label{sec:pipelines}

In this section, we provide an overview of pipelines for Open \gls{ran} automation, and showcase an example of pipeline for the \gls{ci}, \gls{cd}, and \gls{ct} of Open \gls{ran} \glspl{gnb}.

\gls{ci}, \gls{cd}, and \gls{ct} pipelines automatically perform tasks on the \openran infrastructure.
We leverage Tekton---a framework to create \gls{ci}/\gls{cd} workflows for on-premise and cloud system---to implement these tasks and automate our pipelines.
Similarly, we rely on ArgoCD---that implements the GitOps declarative philosophy in a \gls{k8s} microservices cluster---to synchronize and deploy configurations for the host machines and for the tasks from a version-controlled remote repository, ensuring accountability, repeatability, and rollbacks.
The tasks we designed carry out operations such as building container images and deploying them on the physical infrastructure, matching workload resources to the nodes best fit to them (e.g., low-latency nodes), discovering available radio and spectrum bands, performing automatic testing, and monitoring the overall health of system and workloads.

A high-level diagram of a \cicdct pipeline for \openran is shown in Figure~\ref{fig:pipeline-diagram}.
The tasks of this pipeline are divided in four main groups: (i)~test specification parsing; (ii)~repeatable \gls{ran} builds; (iii)~automatic testing of \openran components, and (iv)~test support services.
In the \textit{test specification parsing} step, the details and specifications of the tests to execute are sent to \name. This is the only step that requires some form of human interaction as it involves gathering the test details provided as input, e.g., the \gls{cu}/\gls{du}/\gls{ru} to test, the test requirements, and the relevant technical specifications.
The intent specified in this way is then automatically parsed and converted into services and tasks that can be executed by the compute nodes.
After this initial setup has been carried out, the remaining steps of \name execute in a fully automated manner.

The \textit{repeatable \gls{ran} builds} workflow packages the source code of the component to test (e.g., the \gls{gnb}) into a container image to be instantiated and tested.
First, the source code of the component under test is pulled from a version-controlled repository on a remote git server.
Depending on specific tests, this step can also apply new developed functionalities (e.g., new schedulers or fixes to test) to the source code.
Then, this workflow is used to build a container image that will be stored in an image registry and deployed as workload on the compute nodes.

The \textit{automatic testing} workflow validates the functionalities of the built components.
After being built and stored in the image registry, the new container image is deployed on one of the compute nodes, depending on resource availability and requirements. As these nodes may need to timely interface with the radio devices (e.g., in the case of a \gls{gnb}), they have been optimized for low-latency operations.
Resources mapped to this deployment include radio and spectrum portions---discovered through dedicated services---as well as computational, memory, and networking resources (e.g., high-speed network interfaces to communicate with the radios).
After the resource mapping has been completed, a new container is \textit{deployed} on the compute nodes, and the required resources are allocated to it.
Finally, \textit{testing support services} include services to monitor the health of the host machines and of the deployed workloads, recover them from potential failures or issues, analyze test results, interface with the git, and host a Docker registry for the built images.

\subsection{CI, CD, and CT of Open RAN gNBs}
\label{sec:pipeline-example}

\begin{figure}[t]
    \centering
    \includegraphics[width=\columnwidth]{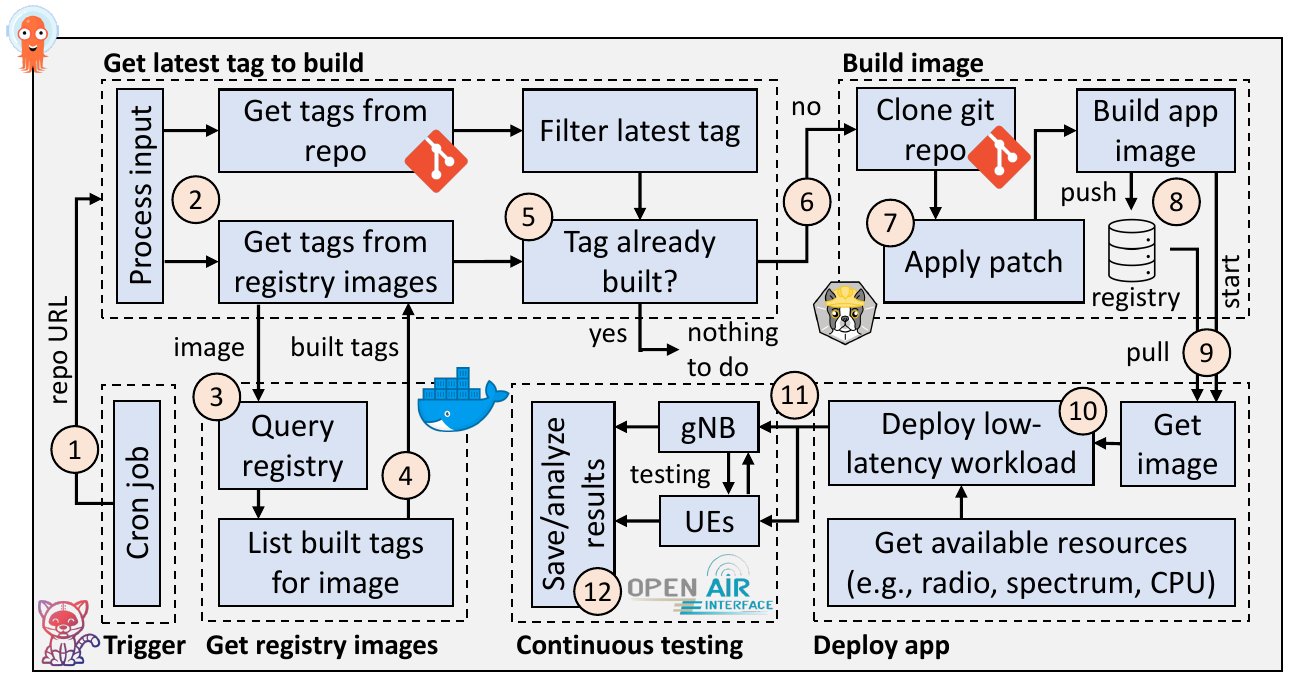}
    \caption{Example of pipeline for CI/CD/CT of Open RAN gNB.}
    \label{fig:pipeline-example}
\end{figure}

An example of a \name pipeline to automatically build and test the functionalities of an Open \gls{ran} \gls{gnb} is shown in Figure~\ref{fig:pipeline-example}.
This pipeline---implemented through Tekton and synchronized with a version-controlled repository through ArgoCD---consists of six main tasks, shown in the figure with a dashed line:
(i)~periodic triggers;
(ii)~get latest tag to build;
(iii)~get registry image;
(iv)~build image;
(v)~deploy app,
and (vi)~continuous testing.

The \textit{periodic triggers} task runs a \textit{cron job} that starts a \textit{test}, whose \textit{specification} was parsed a priori (Figure~\ref{fig:pipeline-diagram}), and executes the \textit{get latest tag to build} job to check changes on a git repository (Step~1 in Figure~\ref{fig:pipeline-example}).
This task receives as input the URL of the repository to be monitored and checks the latest tag published on it (Step~2).
Then, during the \textit{get registry images} task, it queries the image registry (Step~3), part of the \textit{test support services}, for the tags built for the Docker image of the Open \gls{ran} software to test (Step~4).
The list of tags is compared with the latest tag available on the git repository (Step~5).
If no new tags have been released, the pipeline ends; otherwise, the task to \textit{build} the new \textit{image} is started (Step~6) to realize a \textit{repeatable \gls{ran} build}.
The goal of steps~1-6 is to decide whether or not a new container image needs to be built.
These steps take $11$\:s to execute, on average.

The image build task aims at realizing a \textit{repeatable \gls{ran} build} from targeted source code, and is carried out through Buildah, an open-source \cicd tool to build containers compliant with the \gls{oci} specifications---and that can, thus, be deployed on a variety of platforms including Docker, \gls{k8s}, and OpenShift.
We set up Buildah to execute a multi-stage build that first clones the git branch/tag to build, optionally applies patches to the source code (Step~7), builds application and required dependencies, and transfers the built executable to the final image, which is saved in the image registry (Step~8).
Since the build process tailors the final executable to the CPU architectures of the compute nodes, this process is executed on the low-latency nodes where the workload will be deployed.
For instance, building a \gls{gnb} image takes $20$\:minutes, on average, in our setup.

The newly built image is then pulled from the registry (Step~9), matched with the available host machine resources (e.g., radio resources, spectrum availability, CPUs, RAM, physical interfaces, etc.), and \textit{deployed as a new application container} on the low-latency nodes (Step~10, potentially stopping an older running instance) according to the \textit{automatic testing} operations of Figure~\ref{fig:pipeline-diagram}.
As an example, stopping a previously deployed \gls{gnb} container, pulling the updated image from the registry, and deploying it on our system takes $58$\:s, on average, $34$\:s of which are required to terminate the previous \gls{gnb} instance and release the resources used by it.

As part of the automatic testing, the deployed container is used for the \textit{continuous testing} of Open \gls{ran} functionalities (Step~11).
In the case of \gls{gnb} testing, this is done by connecting commercial \gls{ue} devices and exchanging traffic generated via benchmarking tools such as iPerf.
We implemented \glspl{ue} through Sierra Wireless EM9191 5G modems connected to Intel NUC computers, and tested in the \gls{cbrs} band.
Upon completion of the tests, relevant \glspl{kpi} are stored in a database for later analysis and visualization (Step~12) performed by the \textit{test support services}.
Finally, it is worth noticing that, while we showcased the above pipeline as an end-to-end workflow, the single tasks can also be run independently, e.g., to only build a novel container image, or to only test its functionalities.

\section{\name OpenShift-based Automated Infrastructure and Microservices}

The pipelines described so far can be implemented as part of a larger container platform, such as \gls{k8s}, OpenShift, or OKD.
These orchestrators offer a flexible virtualization environment to instantiate and manage workloads in the form of containerized applications---or \textit{pods}---based on microservices, and manage their lifecycle, including requirements for networking, storage, and replication, among others. 

\begin{figure}[t]
    \centering
    \includegraphics[width=\columnwidth]{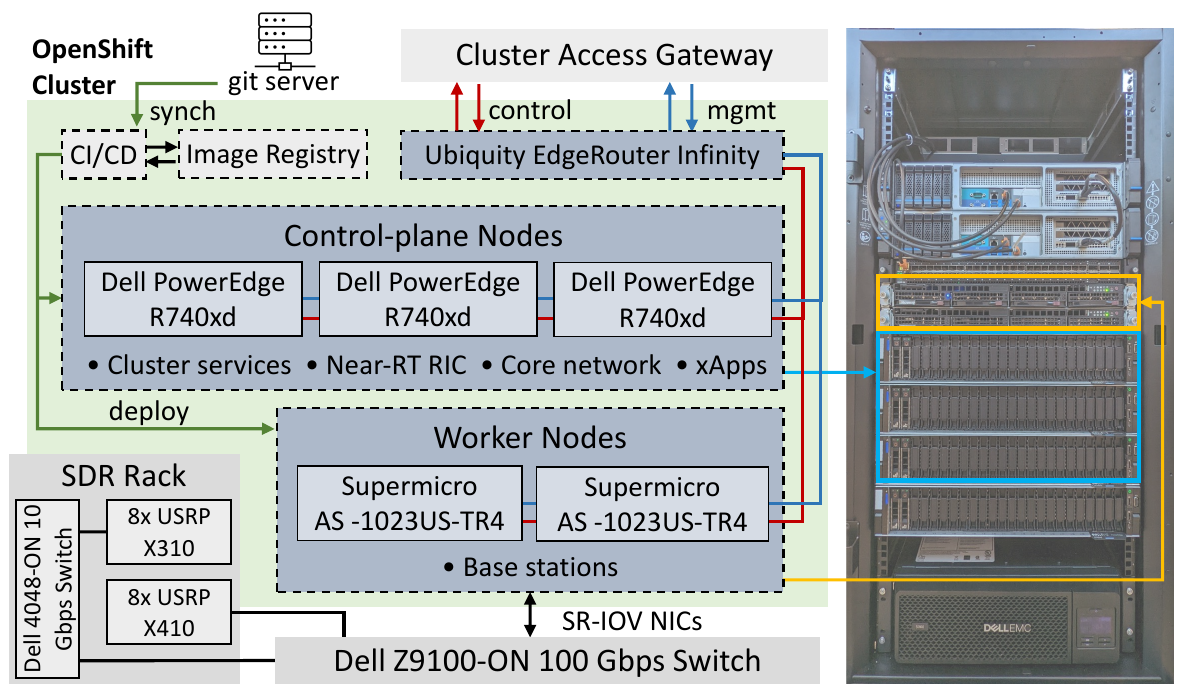}
    \caption{\name architecture.}
    \label{fig:architecture}
\end{figure}

We built \name on top of Red Hat OpenShift Container Platform, which abstracts the system complexity (e.g., in terms of heterogeneous nodes and compute capabilities, different CPUs, \glspl{nic}, RAM) through high-level configuration files.
The architecture of \name is depicted in Figure~\ref{fig:architecture}.
At a high-level, it entails an OpenShift-based cluster with three control-plane nodes (Dell PowerEdge R740xd with Intel Xeon CPUs with 32 logical cores and $192$\:GB RAM) and two worker nodes (Supermicro AS-1023US-TR4 with AMD EPYC CPUs with 32 logical cores and $256$\:GB RAM), and networking infrastructure.
The latter includes a $100$\:Gbps Dell EMC Z9100-ON switch that connects 8~USRP X410 \glspl{sdr} to the low-latency nodes via their \gls{sriov}-enabled \glspl{nic} and QSFP28 cables.
In addition, a $10$\:Gbps Dell EMC 4048-ON switch connects to 8~USRP X310 \glspl{sdr} via SFP+ cables.
Finally, a link aggregation group bounds the two switches through two $40$\:Gbps interfaces, thus allowing the low-latency nodes to communicate with the USRP X310 \glspl{sdr} as well.
Overall, these \glspl{sdr} connect to antennas deployed across a $2240$~square feet indoor office space representative of a private 5G deployment, and they are able to operate in the sub-$7.2$\:GHz \gls{rf} spectrum~\cite{bertizzolo2020arena}.

Control-plane nodes run most of the services required to manage the cluster, as well as generic workloads.
These include \gls{osc} near-RT \gls{ric}, and core networks from Open5GS and A5G Networks.
Worker nodes, instead, are dedicated to specialized workloads that require low-latency operations, e.g., to interface with radio devices in the case of \glspl{gnb}.
Because of this requirement, the configuration of these nodes requires fine-tuning to enable such operations.

A sample configuration for the low-latency worker nodes is shown in Listing~\ref{lst:performanceprofile-yaml}.
\begin{lstlisting}[float=t,floatplacement=h,language=yaml,style=mystyle-yaml,
caption={Configuration example for \texttt{PerformanceProfile} object.},
label={lst:performanceprofile-yaml}]
apiVersion: performance.openshift.io/v2
kind: PerformanceProfile
metadata:
 name: worker-rt-performanceprofile
status:
  runtimeClass: performance-network-latency
spec:
  workloadHints:
      highPowerConsumption: true 
      realTime: true
  additionalKernelArgs:
    - mitigations=off
    - pci=realloc
    - numa_balancing=enable
    - transparent_hugepage=never
    - skew_tick=1
  cpu:
    reserved: 0,1
    isolated: 2-31
  hugepages:
    defaultHugepagesSize: "1G"
    pages:
    - size: "1G"
      count: 64
  nodeSelector:
     node-role.kubernetes.io/worker-rt: ""
  machineConfigPoolSelector:
     machineconfiguration.openshift.io/role: worker-rt
\end{lstlisting}
These nodes are optimized for minimal latency by disabling energy consumption optimizations in favor of maximum performance (lines~8-10), with additional kernel parameters passed in lines~11-16.
Two logical CPU cores, one for each CPU socket, are \textit{reserved} to run the OpenShift services, while the remaining 30~cores are \textit{isolated} and only used to run user workloads (lines~17-19).
Additionally, portions of the physical memory of these nodes are reserved through the use of huge pages (64~huge pages of $1$\:GB each) to increase the performance of the nodes (lines~20-24). 
Finally, worker nodes leverage dedicated NVIDIA Mellanox ConnectX-6 \glspl{nic}---passed to the pods through \gls{sriov} for a trade-off between latency and high-availability, e.g., to share the same physical interface among multiple pods---to connect to \glspl{sdr} via the Z9100-ON switch (see Figure~\ref{fig:architecture}).
OpenShift also allows clusters to integrate and manage hardware acceleration components to perform \textit{look-aside} or \textit{inline} layer~1 data acceleration.
We plan to integrate GPU-acceleration in \name to offload layer~1 functionalities of the cellular stack onto these units.

The functionalities of the cluster can also be extended through custom microservices to integrate non-standard hardware components.
An example of services auxiliary to the Open \gls{ran} ecosystem implemented by the cluster includes radio discovery functionalities that leverage Flask \glspl{api} and the \texttt{usrp\_find\_devices} UHD routine.
Before deploying workloads that need to interface with the \glspl{sdr}, \name pipelines make an \gls{api} call to the Flask endpoint of this service, which returns the list of available USRP radios discovered through the \texttt{usrp\_find\_devices} utility.
As the pod where this service runs does not need timely communication with the radio devices, it interfaces with them through the use of MacVLAN instead of \gls{sriov}.

Other microservices we implemented on the OpenShift cluster include: (i)~services to allocate workloads on nodes that best fit their requirements (e.g., compute, latency, networking); (ii)~an image registry to store the Docker images that will be then instantiated as workloads, (iii)~and a \cicd automation for the configuration of the cluster nodes, as well as for instantiating \gls{ran} services.
Similarly to the described pipelines, also these routines are configured and performed through a \cicd automation implemented through ArgoCD and Tekton.

\ifexttikz
    \tikzsetnextfilename{throughput}
\fi
\begin{figure*}[t]
\setlength\abovecaptionskip{0pt}
    \centering
    \setlength\fwidth{0.9\textwidth}
    \setlength\fheight{.4\columnwidth}
    \includegraphics[width=\textwidth]{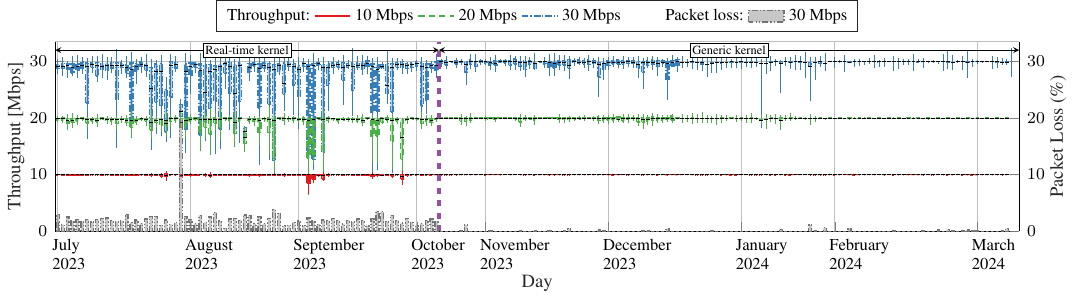}
    \caption{Time evolution of throughput (box plots) and packet loss (bar plot) of an OAI gNB during automated tests performed over 9~months for different data rates. The vertical line in October~2023 marks the switch from real-time to generic kernel of the host machines running the gNB.}
    \label{fig:throughput}
\end{figure*}

\section{Experimental Results}

In this section, we  show results of \name automated tests of an \gls{oai} \gls{gnb}. These tests have been running every 6 hours for approximately 9~months.
At a high-level, the automated testing involves: (i)~instantiating the \gls{oai} \gls{gnb} as an OpenShift pod on one of the low-latency worker nodes connected to the \glspl{sdr} of the testbed; (ii) connecting as \gls{ue} a commercial Sierra Wireless EM9191 5G modem managed by an Intel NUC Mini PC; (iii)~exchanging downlink \gls{udp} traffic at different target data rates---e.g., at $10$, $20$, and $30$\:Mbps in this example---between the \gls{gnb} and the \gls{ue}, and (iv)~reporting the results to dedicated data collectors implemented as additional pods on OpenShift, where the results are also analyzed and stored.

To maintain the \gls{gnb} up-to-date with the latest \gls{oai} releases, \name periodically (i.e., once a week in this example) checks for new \gls{gnb} releases on the \gls{oai} code repository.
If a new release (i.e., git tag) is found, \name builds a new container with the up-to-date \gls{gnb} code, and uses it for all subsequent tests, until a new release is found.
Tests are performed leveraging the \glspl{sdr} and antenna grid of the Arena testbed~\cite{bertizzolo2020arena}, which allows users to run cellular experiments (among others) in an indoor environment representative of a private 5G setup.
The \gls{oai} gNB uses a USRP X410 \gls{sdr} as \gls{rf} front-end, and connects to a commercial 5G core provided by A5G Networks.
Transmissions happen in the \gls{cbrs} band~n48 ($3.62$\:GHz), in \gls{tdd} mode with $162$~\gls{prb} ($30$\:MHz).

Tests are triggered through a call to the Flask \glspl{api} that we added to the \gls{gnb} pod.
Once this \gls{api} call is received, the pod starts the \gls{oai} \gls{gnb}, waits for it to be running, and makes a call to the Intel NUC connected to the Sierra Wireless 5G modem that acts as \gls{ue}.
At this point, the Intel NUC turns on the 5G modem, and waits for it to connect to the \gls{oai} \gls{gnb} and to receive an IP address from the 5G core.
Once connected, the Intel NUC defaults its network routes toward the 5G core (via the \gls{gnb}), and starts an iPerf Docker container, which connects to an iPerf server deployed as a standalone pod on OpenShift, to start the performance tests at the selected rates.
After the iPerf tests terminate, the results gathered by the \gls{ue} are sent to a data-collector service---implemented through a replica of 3~OpenShift pods for redundancy---leveraging the same connection with the \gls{gnb} used for the tests.
Results are analyzed by this service, which extracts relevant metrics and stores them in a dedicated \gls{nfs} volume on OpenShift. They are made accessible through an \texttt{httpd} web dashboard implemented via a replica of 3~OpenShift pods.
Overall, this example of \name automated tests involves a total of 35~OpenShift pods (1~for the \gls{gnb}, 1~for the iPerf server, 3~for the data-collector service, 3~for the dashboard, and 27~for the core), while an end-to-end cellular network can be instantiated in tens of seconds, as demonstrated in~\cite{bonati2023neutran}.

Figure~\ref{fig:throughput} shows the overall throughput achieved by \name automated tests for the three configurations considered.
We show the day in which the experiment was performed on the x-axis, and box plots of the throughput averaged over the 4~daily experiments on the y-axis for each configuration.
Some boxes are missing because of unsuccessful tests (e.g., miscommunication between the \gls{gnb} and the \gls{sdr}, issues with the iPerf server).
The figure also shows the packet loss of the $30$\:Mbps configuration (bar plot).
We notice that, in general, the throughput of the $10$ and $20$\:Mbps configurations (shown in red and green in the figure, respectively) is steady throughout all tests, with the $20$\:Mbps configuration seldom having unstable behavior.
The throughput of the $30$\:Mbps configuration (shown in blue), instead, exhibits performance instability, which is substantially mitigated after updating the kernel configuration of the worker nodes of the cluster (vertical line in October~2023).
Indeed, changing the type of Linux kernel and reducing the CPU performance spikes (i.e., passing the kernel option \texttt{skew\_tick=1}), reduces the performance instability, especially in the $30$\:Mbps case.
Thus, tests before updating this configuration show some instability---due to a combination of issues in the \gls{oai} code (e.g., we found that some releases had issues in the communication among the \gls{gnb} and the USRP \gls{sdr}), interference, and kernel settings---that seems to have been resolved in tests with the updated setup.
This is consistent with the packet loss, only shown for the $30$\:Mbps tests (worst case) in the interest of visualization clarity.

\begin{figure}[t]
\setlength\abovecaptionskip{0pt}
    \centering
    \setlength\fwidth{0.875\columnwidth}
    \setlength\fheight{.32\columnwidth}
    \includegraphics[width=\columnwidth]{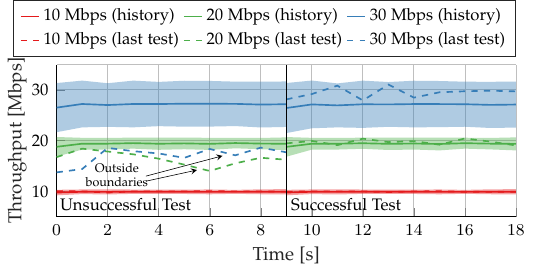}
    \caption{Comparison between unsuccessful and successful tests of an OAI gNB for different targeted data rates. The solid lines show the history of previous tests, with shaded areas depicting their confidence intervals. The dashed lines show the most recent test to compare.}
    \label{fig:build-comparison}
\end{figure}

Figure~\ref{fig:build-comparison} shows the outcome of unsuccessful and successful tests of an \gls{oai} gNB, by comparing them with historical test data collected through \name that provide reference performance levels.
\ifexttikz
    \tikzsetnextfilename{build-comparison}
\fi
The  history of the throughput of previous tests (which serves as a baseline) is shown in solid lines, with the shaded areas representing the variance of the tests. The latest test to compare with the result history is shown with dashed lines.
After performing the automated tests, \name compares the performance of the most recent test with the test history for the same configuration.
If the performance of each test falls within the shaded areas of the test history for the same configuration, the test is marked as successful (e.g., test of Figure~\ref{fig:build-comparison}, right).
Otherwise, the test is marked as unsuccessful (Figure~\ref{fig:build-comparison}, left).
By performing such comparison, \name can flag tests as \textit{passed} or \textit{failed,} which will lead to automatically generated reports in future extensions.

\section{Conclusions}

We introduced \name, an automation framework based on Red Hat OpenShift that leverages the GitOps paradigm to automatically deploy and test softwarized end-to-end 5G and O-RAN-compliant systems in a matter of seconds.
By extending cloud-native \gls{ci}/\gls{cd} pipelines to the \gls{ran}, \name effectively addresses the increasing complexity of operations such as spectrum and radio management, and allocation of heterogeneous resources, devices and distributed components, thus providing the much needed automation for the cellular ecosystem. In this way, \name has the potential to increase the reliability, robustness, and security of software for Open \gls{ran}.
We integrated \name with an over-the-air \gls{sdr} testbed, and demonstrated how it can be used to test open-source protocol stacks for cellular networks, including \gls{oai}, though automated \gls{ct} experiments spanning several~months.

\footnotesize
\bibliographystyle{IEEEtran}
\bibliography{biblio.bib}

\vspace{-1.3cm}
\begin{IEEEbiographynophoto}{Leonardo Bonati} [M'23] is an Associate Research Scientist at the Institute for the Wireless Internet of Things, Northeastern University, Boston. He received a Ph.D.\ degree in Computer Engineering from Northeastern University in 2022. His research focuses on softwarized approaches for the Open RAN of next generation of cellular networks.
\end{IEEEbiographynophoto}

\vspace{-1.3cm}
\begin{IEEEbiographynophoto}{Michele Polese}
[M'20] is a Research Assistant Professor at the Institute for the Wireless Internet of Things, Northeastern University, Boston. He received his Ph.D.\ degree from the University of Padova in 2020. He then joined Northeastern University as a research scientist. His research interests are in the analysis and development of protocols and architectures for future generations of cellular networks.
\end{IEEEbiographynophoto}

\vspace{-1.3cm}
\begin{IEEEbiographynophoto}
{Salvatore D'Oro} [M'17] is a Research Assistant Professor at the Institute for the Wireless Internet of Things, Northeastern University, Boston. He received his Ph.D.\ degree from the University of Catania in 2015. His research interests include optimization, artificial intelligence, and security applied to Open RAN systems.
\end{IEEEbiographynophoto}

\vspace{-1.3cm}
\begin{IEEEbiographynophoto}{Pietro Brach del Prever}
[S'23] is a Ph.D.\ student at the Institute for the Wireless Internet of Things, Northeastern University, Boston. He received his M.S.\ degree in Mechatronic Engineering from the Polytechnic University of Turin, Italy, in 2022. His research interests focus on NextG wireless systems.
\end{IEEEbiographynophoto}

\vspace{-1.3cm}
\begin{IEEEbiographynophoto}{Tommaso Melodia}
[F’18] received a Ph.D.\ in Electrical and Computer Engineering from the Georgia Institute of Technology in 2007. He is the William Lincoln Smith Professor at Northeastern University, the Director of the Institute for the Wireless Internet of Things, and the Director of Research for the PAWR Project Office. His research focuses on wireless networked systems.
\end{IEEEbiographynophoto}

\end{document}